\documentclass[a4paper,11pt]{article}
\usepackage[english]{babel}
\usepackage[utf8]{inputenc}
\usepackage{siunitx}
\usepackage[version=4]{mhchem}
\usepackage{graphicx}
\usepackage{caption}
\usepackage{subcaption}
\usepackage{cite}
\usepackage{geometry}
\usepackage[skip=2pt,font=footnotesize,margin=1cm]{caption}

\begin{document}
\begin{center}
\textbf{\Large Mid-infrared homodyne balanced detector for quantum light characterization} \\
\vspace{5mm}

Tecla Gabbrielli\textsuperscript{1,2,*},
Francesco Cappelli\textsuperscript{1,2},
Natalia Bruno\textsuperscript{1,2},
Nicola Corrias\textsuperscript{1,2},
Simone Borri\textsuperscript{1,2,3},
Paolo De Natale\textsuperscript{1,2,3},
Alessandro Zavatta\textsuperscript{1,2,3}
\\

\bigskip
\textit{\small\textsuperscript{1} Istituto Nazionale di Ottica (CNR-INO), Largo Enrico Fermi 6, 50125 Florence, Italy}
\\
\vspace{1mm}
\textit{\small\textsuperscript{2} European Laboratory for Non-linear Spectroscopy (LENS), Via nello Carrara 1, 50019 Sesto Fiorentino, Florence, Italy}
\\
\vspace{1mm}
\textit{\small\textsuperscript{3}Istituto Nazionale di Fisica Nucleare (INFN), Sezione di Firenze, 50019 Sesto Fiorentino, Florence, Italy}

\bigskip
* gabbrielli@lens.unifi.it
\end{center}

\begin{abstract}
We present the characterization of a novel balanced homodyne detector operating in the mid-infrared. The challenging task of revealing non-classicality in mid-infrared light, e.~g. in quantum cascade lasers emission, requires a high-performance detection system. Through the intensity noise power spectral density analysis of the differential signal coming from the incident radiation, we show that our setup is shot-noise limited. We discuss the experimental results with a view to possible applications to quantum technologies, such as free-space quantum communication.  
\end{abstract}

\section{Introduction}

Balanced homodyne detection is an effective measurement technique, widely used by the quantum optics community~\cite{yuen1983noise,Shapiro:1985,raymer1995ultrafast,Lambrecht_1996,loudon:2000quantum,zavatta2002time,sasaki2006multimode,kumar2012versatile}, as it allows the reconstruction of quantum states of light by retrieving their field quadratures. It is based on a differential measurement carried out after mixing the signal of interest with a reference radiation, named Local Oscillator (LO), on a 50/50 beam splitter followed by two identical detectors.
This technique has all the advantages of the balanced detection, where the common noise (e.~g. correlated noise due to the photon-generation process or amplification) is suppressed by measuring the difference between the two balanced parts of the optical beam.
Any possible classically-correlated contribution affecting the measurement is, therefore, cancelled out increasing the detection sensitivity as required for reaching the standard quantum limit~\cite{Schumaker:84}. Investigation of quantum light through balanced homodyne detection has been extensively pursued in the near-infrared. As a matter of fact, this is the region where the first efficient systems have been developed~\cite{raymer1995ultrafast,Lambrecht_1996,loudon:2000quantum,zavatta2002time} and utilized~\cite{costanzo2017measurement,Zavatta:2020}. Besides, the technological progress of near-infrared components goes hand-in-hand with the worldwide demand for communications devices at telecom wavelengths~\cite{kaiser2016fully,mondain2019chip}.
In this framework, balanced homodyne detection is a useful tool, widely exploited for continuous-variable quantum communication both in optical fiber and free-space links~\cite{Ralph:1999,Semenov:2009,Elser:2009}. 
The mid-infrared (MIR) spectral region ($\lambda > \SI{3}{\micro \meter}$) is a promising alternative to the near-infrared for free-space-optical communication~\cite{Temporao}. In fact, considering the well-reduced Rayleigh scattering cross section compared to the visible/near-infrared, the atmosphere's MIR transparency window between 3 and \SI{5}{\micro m} makes MIR radiation an excellent candidate for free-space communication applications. Up to now, MIR light has been widely investigated and employed for spectroscopy applications. Here indeed many molecules of atmospheric and astrophysical interest can be investigated on their strongest ro-vibrational transitions~\cite{wysocki:2005,lee:2007,Bartalini:2009,Galli:2013a,Galli:2013b,Galli:2014a,Galli:2016b,Coddington:2016,Campo:2017,Consolino:2018,Borri:2019a,Picque:2019,Karlovets:2020}. Also in this field, the availability of quantum MIR sources can set the scene for compelling quantum sensing applications~\cite{RevModPhys.Degen:2017}.

The goal of this work is to explore the extension of quantum balanced homodyne detection to the MIR spectral region, by demonstrating novel technologies for fully exploiting the advantages to operate in a quantum regime. Balanced detection has already been investigated in the MIR for classical applications such as frequency-modulation spectroscopy~\cite{carlisle:1989}, difference-frequency laser spectroscopy~\cite{chen:1998}, balanced radiometric detection~\cite{sonnenfroh:2001}, and Doppler-free spectroscopy~\cite{Bartalini:09}. Other optical schemes suitable for single-photon quantum applications, such as coincidence measurements~\cite{mancinelli2017} or free-space Quantum Key Distribution with discrete variables~\cite{aellen2008feasibility}, have so far been studied. In this work, we evaluate the possibility of investigating continuous-variable quantum physics in the MIR through a novel Balanced Homodyne Detector (BHD).  
In particular, our BHD has been tested with Quantum Cascade Lasers (QCLs), chip-scale semiconductor-heterostructure devices based on intersubband transitions in quantum wells, operating in the mid-to-far infrared~\cite{Faist:1994,Tombez:2013a}.
The development of our BHD allows the investigation of the quantum properties on QCLs radiation which are yet unexplored. Broadband QCLs can emit frequency combs due to the high third-order non-linearity which characterizes the active region and enables a Four-Wave Mixing (FWM) parametric process in their waveguide~\cite{hugi:2012,Friedli:2013,Riedi:2015,Burghoff:2014,Faist:2016,Cappelli:2016,CappelliConsolino:2019,Mezzapesa:2019,Consolino:2020}. From a quantum optical point of view, FWM makes QCLs potential non-classical state emitters. Indeed, the possibility of engineering squeezed and color entangled states via FWM has been already demonstrated in several optical systems~\cite{levenson:1985,slusher:1985squeezing,mccormick:2007,Dutt:2015}. 

In the following sections we show that the novel BHD here presented is shot-noise-limited and suitable for directly unveiling non-classicality in MIR light. This represents the first experimental step for the investigation and exploitation of non-classical correlations in the light emitted by QCLs.

\section{Methods}\label{sec:Theory of balanced detection}
\subsection{Theory of balanced detection} 
\label{subsec:theory}

We describe the theory of BHD, composed of a 50/50 beam splitter and two detectors~\cite{loudon:2000quantum}, to test whether a given detector can operate at the shot-noise level when a single-mode radiation is used as the LO. Typically, the LO is assumed to be a coherent state with a well-defined photon-number variance, equal to the mean number of emitted photons~\cite{loudon:2000quantum}. In the following description, we consider the incident radiation to have an arbitrary variance $(\Delta n_R )^2$ to take into account the extra noise that is present in our QCL sources (see section~\ref{sec:Results and discussion}). 
\\In a real setup, optical signals are affected by losses caused by absorption and reflections from optical components. In Quantum Optics theory, losses can be represented as a beam splitter that couples the radiation with the vacuum, characterized by the coefficients $R=i(\sqrt{1-\eta_1})$ and $T= \sqrt{\eta_1}$ where $\eta_1$ is the overall optical transmission efficiency, taking into account any attenuation due to the different components of the experimental setup~\cite{loudon:2000quantum}. The optical losses budget should be carefully addressed according to the specific application of the balanced detector. In the case of homodyne detection, the LO acts as the reference radiation and the relevant optical losses are the ones affecting the signal of interest mixed via the beam splitter with the LO~\cite{loudon:2000quantum}. On the contrary, when the balanced detector is used for characterizing the statistics of the laser source employed as LO, the LO becomes the radiation under study and, therefore, the optical losses affecting it become relevant. The two different scenarios are discussed in the corresponding experimental context in section~\ref{sec:ExpSetup}.
\begin{figure}[!htbp]
    \centering
    \includegraphics[width=0.7\columnwidth]{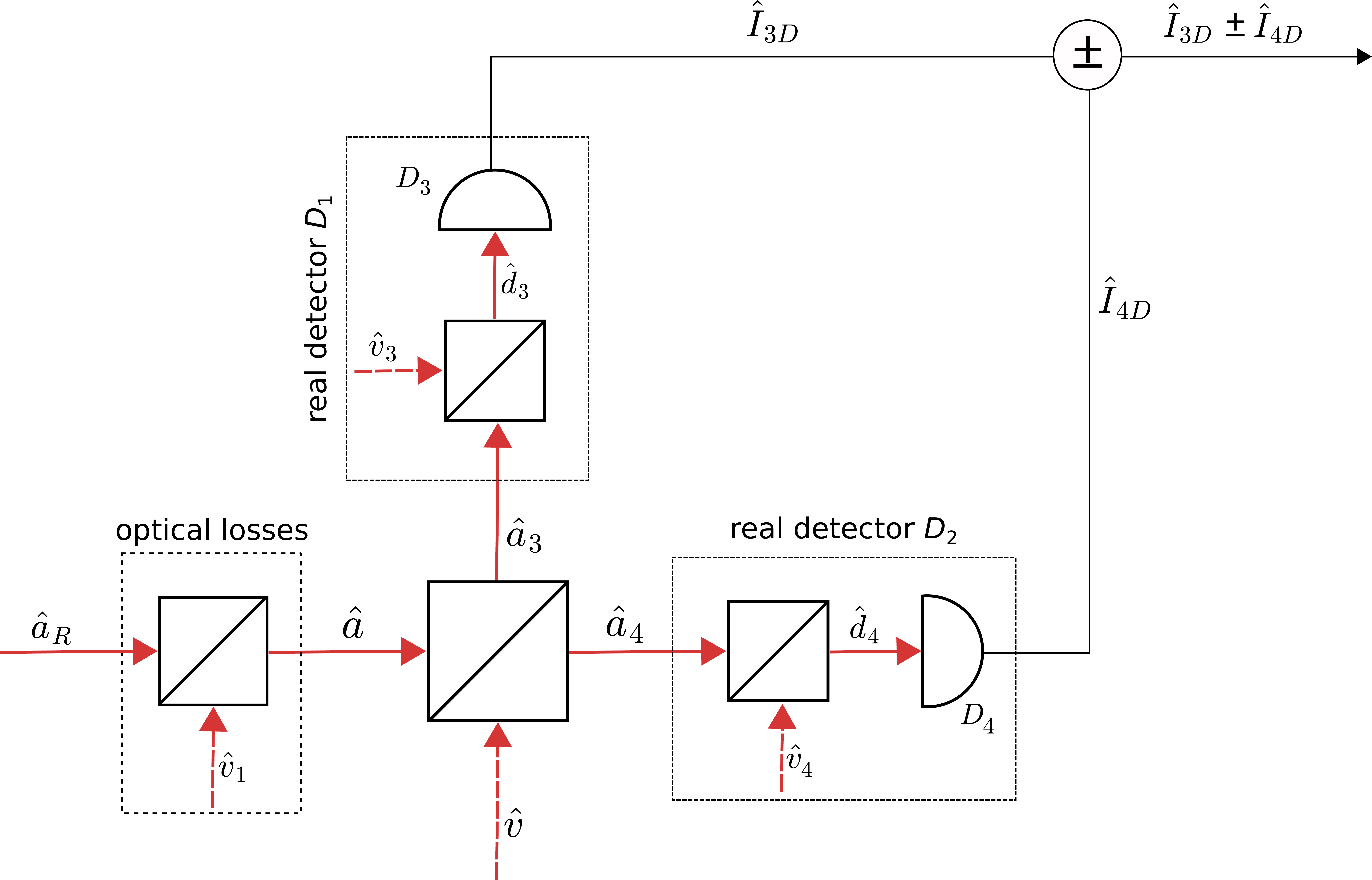}
    \caption{
Scheme of the sum and difference measurement, where an attenuation $1-\eta_1$ is placed before the 50/50 beam splitter and the two real detectors are considered with quantum efficiency $\eta_{qe}$. In the scheme, the fields are represented by the annihilation operators. In particular: $\hat{a}_R$ represents the LO; $\hat{v}$,$\hat{v}_1$,$\hat{v}_3$, $\hat{v}_4$ represent the vacuum entering in the different beam splitters; $\hat{a}$,  $\hat{a}_3$,  $\hat{a}_4$,  $\hat{d}_3$,  $\hat{d}_4$ are the fields at the output of the corresponding beam splitter, resulting from the mixing of the incident fields~\cite{loudon:2000quantum}; $D_3$ and $D_4$ are the ideal detectors that, together with the respective beam splitter describing the quantum efficiency, constitute the corresponding real detectors $D_1$ and $D_2$.}
    \label{fig:sommadiffreale}
\end{figure}
In practice, also the two detectors have losses, resulting in a ratio between flowing electrons and number of incident photons lower than one (quantum efficiency $\eta_{qe} < 100\%$). These losses can be modelled with a beam splitter as well placed before an ideal detector ($D_3$ and $D_4$, Fig.~\ref{fig:sommadiffreale}). For this model, the real detectors are assumed to be identical (same quantum efficiency), with no saturation, and with an instantaneous and linear responsivity in time. Any time dependence in the creation and annihilation operators is neglected. Furthermore, the detection system is assumed to be perfectly balanced to benefit from the advantages of a balanced detection in term of noise suppression~\cite{Schumaker:84,loudon:2000quantum}. In a setup as the one depicted in Fig.~\ref{fig:sommadiffreale}, it is possible to derive a relation between the real detected quantities (labelled with $D$ in the equations below) and the ones of the incident radiation. The currents at the outputs of the two detectors, $\hat{I}_{3D}$ and $\hat{I}_{4D}$, are proportional to the incident flux of photons onto the corresponding detectors. Therefore, integrating the sum and the difference of the output signals over the measurement time leads to the following results:
\begin{eqnarray}
\langle \hat{N}^D_+ \rangle & = & \eta \langle \hat{n}_R \rangle \label{eq:sum},\\
\left (\Delta {N}^D_+ \right)^2 & = & \eta^2 \left (\Delta {n}_R \right)^2 + \eta (1-\eta)  \langle \hat{n}_R \rangle \label{eq:realvariancesum},\\
\langle \hat{N}^D_- \rangle & = & 0 \, \label{eq:dif},\\
\left(\Delta N^D_- \right)^2 & = & \eta \langle \hat{n}_R \rangle \label{eq:realvariance},
\end{eqnarray}
where $\hat{N}^D_+$ (Eqs.~\ref{eq:sum} and \ref{eq:realvariancesum}) is the sum of the detected photon-number signals, $\hat{N}^D_-$ (Eqs.~\ref{eq:dif} and \ref{eq:realvariance}) is the difference, $\eta = \eta_1 \eta_{qe}$, and $\hat{n}_R$ is the number of photons emitted by the LO source. From this derivation, it is clear that an accurate analysis of the losses is needed: losses reduce the measured signal and add an extra term to the variance of the sum given by the coupling with the vacuum field. An excess of attenuation can lead to a signal lying under the background noise floor of the setup and/or to a measured light statistics dominated by the vacuum fluctuations. In the case of a coherent state $\left (\Delta {n}_R \right)^2 = \langle \hat{n}_R \rangle $, the retrieved sum signal via Eq.~\eqref{eq:realvariancesum} is at the shot-noise level $ \left (\Delta {N}^D_+ \right)^2 =  \eta \langle \hat{n}_R \rangle $ and the vacuum does not alter the measured light statistics. More generally, in a regime not dominated by vacuum fluctuations, by comparing Eq.~\eqref{eq:realvariancesum} with Eq.~\eqref{eq:realvariance}, it is possible to understand whether the statistics of the incident light (LO) is shot-noise limited.
\subsection{Homodyne detector characterization}
As discussed in the previous section, it is possible to characterize a balanced homodyne detection setup by sending only the LO on the beam splitter and measuring the difference signal between the two detector outputs. In the limit of the detector's linear responsivity, the differential noise (Eq.~\eqref{eq:realvariance}) is directly proportional to the incident power of the LO and corresponds to the shot noise. 
Furthermore, the balanced detector can be applied for the characterization of the LO statistics: in principle, the laser intensity noise can be retrieved from the sum of the two detectors output signals (Eq.~\eqref{eq:realvariancesum}). It is possible to understand if the statistics of the LO is shot-noise-limited by comparing the sum (Eq~\eqref{eq:realvariancesum}) with the difference (Eq.~\eqref{eq:realvariance}). The BHD differential background noise, which comprises the dark current of the detectors and the electronic noise, sets the BHD sensitivity limit and determines the minimum measurable noise level. The clearance, given by the noise power ratio between the measured shot noise and the differential background noise, is an important feature for a BHD, as it contributes to the overall detection efficiency~\cite{appel2007electronic}. The maximum clearance of the BHD is obtained for the maximum incident LO power before detector saturation. Working in the linear regime is essential for having a direct link between the current's statistics at the detector output and the photon statistics of the incident radiation and, consequently, to get accurate results. Finally, another crucial point for the test of the BHD is a thorough analysis of the losses. Indeed, as described in Eqs.~\eqref{eq:realvariancesum} and ~\eqref{eq:realvariance}, each loss (e.g. due to optical elements or quantum efficiencies of the detectors) couples the light under analysis with the vacuum field, and thus affects the measured statistics~\cite{loudon:2000quantum}. 
\subsection{Experimental setup}
\label{sec:ExpSetup}
\begin{figure}[!htbp]
    \centering
  \includegraphics[width=0.7\columnwidth]{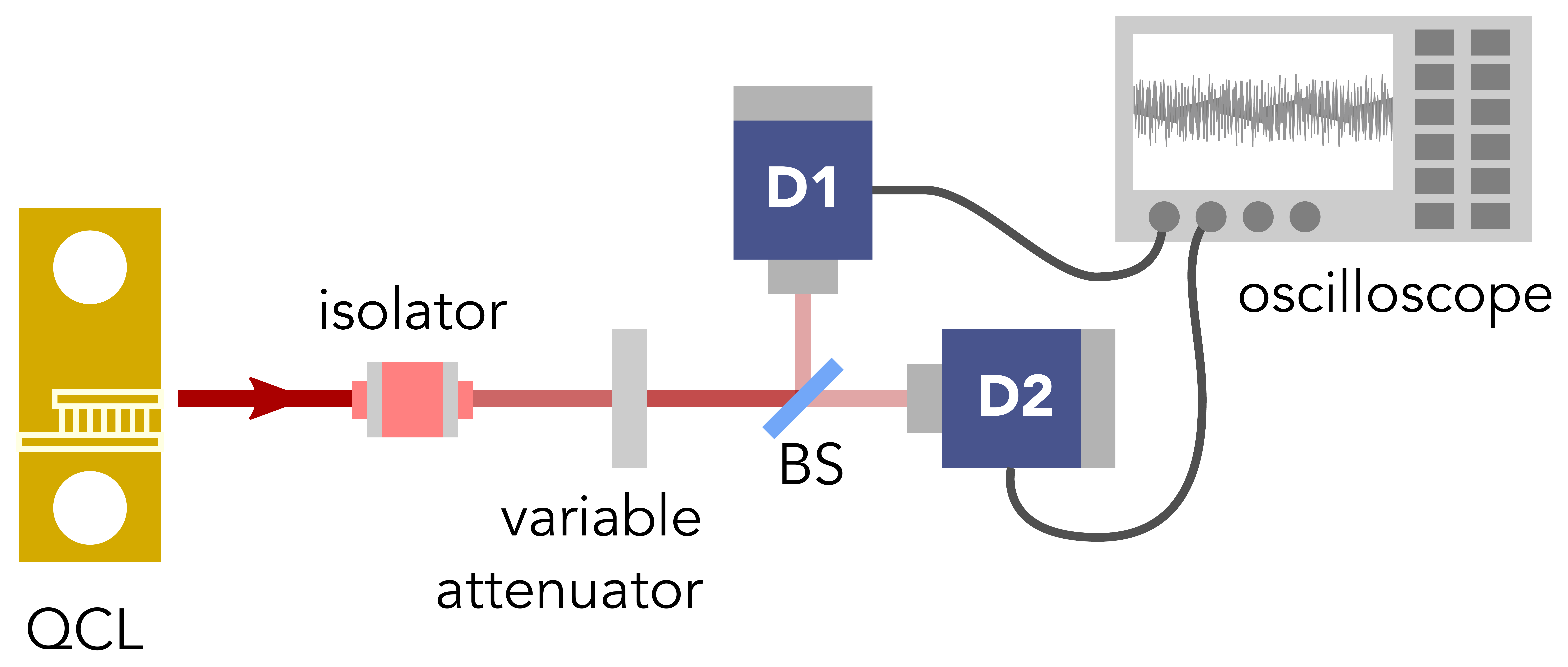}
  \caption{Sketch of the BHD characterization setup. A single-mode QCL is used as LO and is sent on the BHD made of a 50/50 beam splitter (BS) and two \ce{HgCdTe} photovoltaic detectors. After a preamplification stage, the detector output AC-signals are acquired in the time domain using an oscilloscope. To avoid any detector power saturation, a variable attenuator is used to control the incident laser power.}
  \label{fig:50-50setup}
 \end{figure}
The setup used to characterize the BHD is schematically shown in Fig.~\ref{fig:50-50setup}. It is composed of a 50/50 \ce{CaF2} beam splitter, coated for a wavelength range from~\SIrange{2}{8}{\micro\meter}, and two commercial preamplified photovoltaic \ce{HgCdTe} detectors (VIGO, PVI-4TE-5-2x2) characterized by a nominal bandwidth of \SI{180}{\mega\hertz} and a spectral response spanning  from~\SIrange{2.5}{5.0}{\micro \meter} \cite{note3}. These detectors are equipped  with a commercial two-stage preamplifying system (VIGO, MIP-10-250M-F-M4): the first stage is a DC-coupled transimpedance amplifier, the second stage is an AC-coupled amplifier with a measured gain of 26.5 in voltage. The detectors are cooled down to $T=\SI{200}{K}$ by a four-stage-peltier cooling system via a thermoelectric cooler controller (VIGO,PTCC-01-BAS).
The 50/50 splitting is done with a precision $|R|^2-|T|^2= 0.2\%$, calculated via DC signals. The signals at the output of the detectors are acquired in the time domain with a sample rate of \SI{625}{\mega S/s} through two different channels of an oscilloscope with a bandwidth of \SI{200}{\mega \hertz}. The time duration of each acquisition is \SI{1}{\milli \second}.
\begin{figure}[!htbp]
    \centering
  \includegraphics[width=0.8\columnwidth]{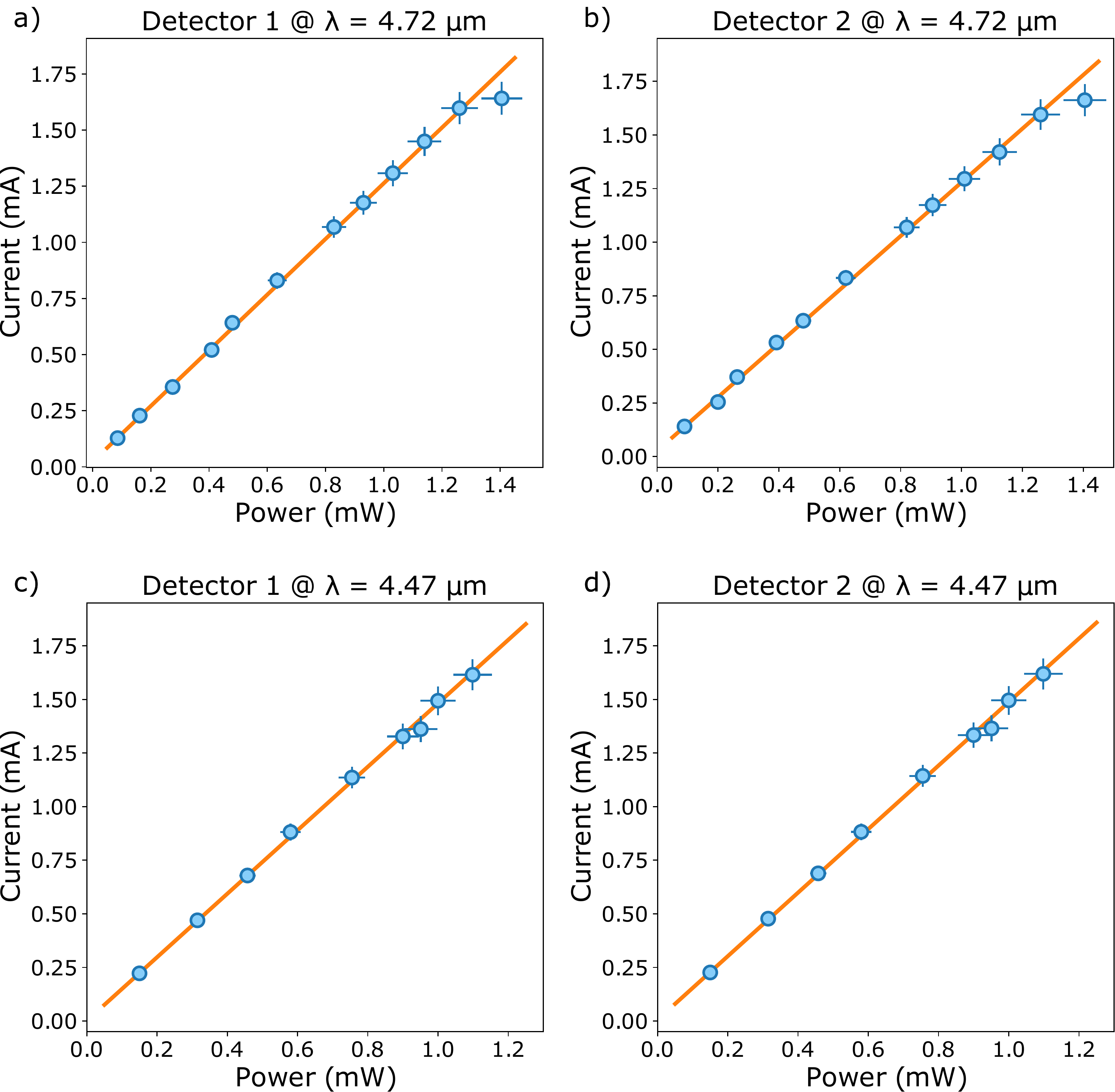}
  \caption{Detectors responsivities measured at \SI{4.72}{\micro \meter} (a,b) and \SI{4.47}{\micro \meter} (c,d) incident radiation. From the linear fit (orange curve) the responsivities of detector 1 and detector 2 are respectively:  $\mathcal{R}_{1}=\SI{1.24 \pm 0.02 }{A/W}$ (a) and $\mathcal{R}_{2}=\SI{1.25 \pm 0.02 }{A/W}$ (b) for $\lambda=\SI{4.72}{\micro \meter}$  incident radiation, $\mathcal{R}_{1}=\SI{1.48 \pm 0.01 }{A/W}$ (c) and $\mathcal{R}_{2}=\SI{1.48 \pm 0.02 }{A/W}$ (d) for $\lambda=\SI{4.47}{\micro \meter}$ incident radiation. For low-power levels, the error bars are covered by the size of the data points (blue circles).
  \label{fig:responsivity}}
 \end{figure}

We tested the BHD with two different continuous-wave single-mode QCLs emitting at $\lambda=\SI{4.47}{\micro \meter}$ and $\lambda=\SI{4.72}{\micro \meter}$. The two lasers are powered by ultra-low-noise current drivers (ppqSense, QubeCL15-P) with a typical current noise density of $\SI{200}{pA/\sqrt{\mathrm{Hz}}}$, to minimize excess technical noise. The QCL radiation first passes through a MIR optical isolator (wavelength working range from \SIrange{4.5}{4.7}{\micro\meter}) and is then attenuated by a variable attenuator used for controlling the LO power impinging on the BHD independently of the laser's operating regime. In the case of maximum transmission through the variable attenuator, taking into account the contribution from all the optical elements (attenuator, isolator, beam splitter, lenses, mirrors), the total optical transmission is \SI{47 \pm 1}{\%} at \SI{4.72}{\micro \meter} and \SI{55 \pm 1}{\%} at \SI{4.47}{\micro \meter}. 

As shown in section~\ref{sec:Theory of balanced detection}, the quantum efficiency is another key parameter to be taken into account in the loss budget. It can be calculated from the responsivities ($\mathcal{R}$), reported in Fig.~\ref{fig:responsivity}, as $\eta_{qe} = \mathcal{R} hc/ (\lambda e)$, where $h$ is the Plank constant, $c$ is the speed of light, and $e$ is the electron charge. This leads to a quantum efficiency of \SI{33 \pm 1}{\%} for both the two detectors at  $\lambda=\SI{4.72}{\micro \meter}$ (Figs~\ref{fig:responsivity} (a) and (b)). At $\lambda=\SI{4.47}{\micro \meter}$ the responsivities of the detectors are higher (Figs.~\ref{fig:responsivity} (c) and (d)), in agreement with the curve reported in the datasheet, which has a maximum at \SI{4.5}{\micro \meter}. The corresponding quantum efficiency is \SI{41 \pm 1}{\%}.
\paragraph{Losses budget for balanced homodyne detection.} For homodyne detection applications of the BHD, we must consider both the quantum efficiency and the optical losses contribution due to the optical components from the beam splitter on. This leads to a detection efficiency up to 40\%. In the MIR, a large contribution to the losses comes from the Fresnel reflection off the optical elements, due to the relatively large refractive index of their materials.
Notice that the quantum efficiency of the \ce{HgCdTe} detectors is not limited by fundamental properties of the material, but mostly by its purity.
Besides, the available anti-reflection coatings for this spectral region are generally less effective and much more expensive than the ones for visible and near-infrared wavelengths, where, thanks to a more advanced technology, higher quantum efficiencies are easier to achieve, as required for advanced measurements such as quantum state tomography~\cite{lvovsky2009continuous}. 
 In addition to the adoption of effective anti-reflection coatings, the detection efficiency can be enhanced by increasing the absorption probability of the light by the photodiode. This can be done,  for example, by placing a golden surface on the back of the semiconductor medium acting as a retroreflector.


\paragraph{Losses budget for laser source characterization.} 
To characterize the laser source that we aim to employ as LO, we can use the presented detector (Fig.\ref{fig:50-50setup}) as a direct balanced detector, in which we mix the light under study with the vacuum and compare the sum with the difference of the two acquired signals, as reported in section~\ref{sec:Theory of balanced detection}. In this scenario, the LO is no longer a reference radiation but it becomes the light under investigation itself. For studying the light emitted by the laser source, both optical losses and quantum efficiency are relevant, as they change the statistics of the measured light field via coupling with the vacuum (Eq.~\eqref{eq:realvariancesum}). With an optical transmissivity of 55\% and a quantum efficiency of 41\%, the total achievable maximum detection efficiency is around 23\%. It is worth noting that this value depends both on the detection system and on the source. Indeed, when the laser output power overcomes the detector saturation level (e.~g. $ P>\SI{1.2}{mW}$, Figs.~\ref{fig:responsivity} (a) and (b)), an attenuator is required. This introduces losses and affects the overall efficiency. Moreover, when using lasers which are sensitive to optical feedback as QCLs, 
an optical isolator is required. This is the case of our setup, as shown in Fig.~\ref{fig:50-50setup}, in which an optical isolator with a transmissivity around 70\% (60\%) at \SI{4.47}{\micro\meter} (\SI{4.72}{\micro\meter}) is used to prevent optical feedback perturbing laser operation. In optimal conditions, that is an emission power below the detector saturation level and no isolator, only the losses due to the remaining optical elements (beam splitter, lenses and detectors) have to be considered and the detection efficiency can increase up to 40\%.

\section{Results and discussion}
\label{sec:Results and discussion}
\begin{figure}[!htbp]
    \centering
  \includegraphics[width=0.99\columnwidth]{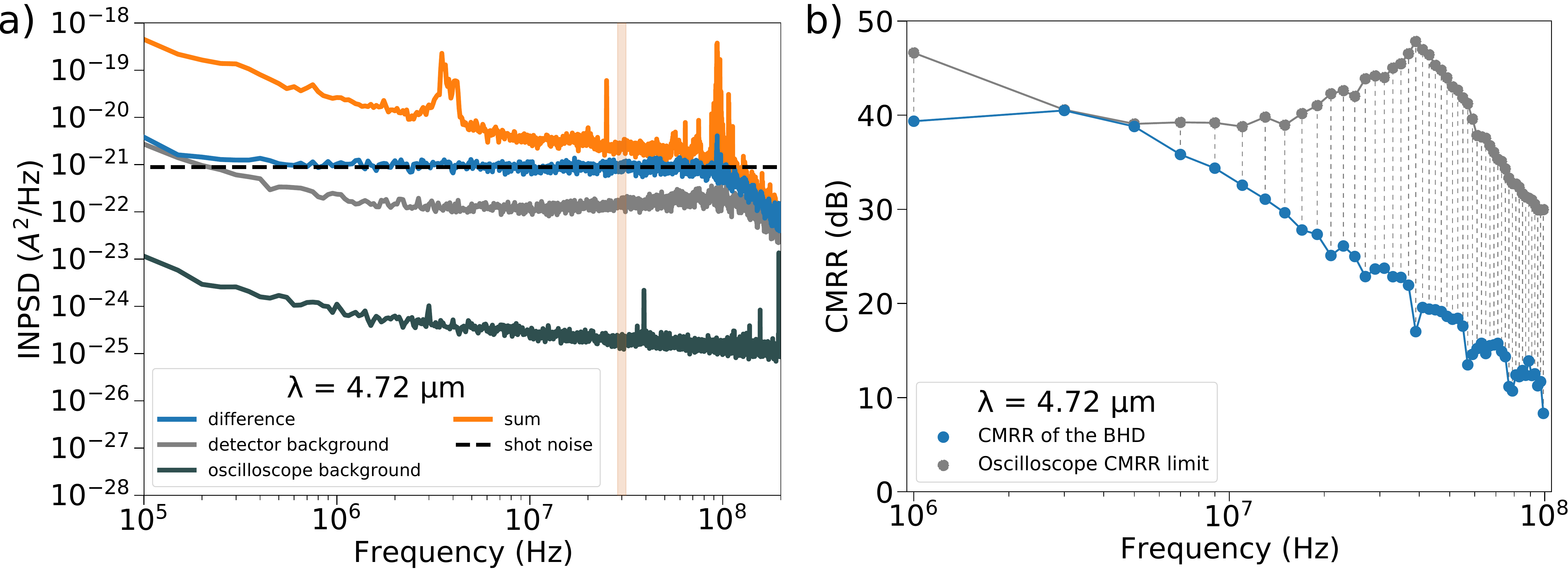}
  \caption{(a) Example of INPSDs of the sum (orange trace) and the difference (blue trace) of the AC signals compared to the difference of the detector background (grey trace) and the difference of the two oscilloscope channels backgrounds (petroleum trace). The dashed black line represents the theoretical one-sided PSD shot-noise level for an ideal detector (i.e. with an infinite bandwidth). For frequencies higher than $\SI{100}{MHz}$, the measured spectra show a drop below the theoretical curve due to the finite detector bandwidth. The vertical orange area shows the frequency window of \SI{3}{MHz} centred at \SI{30}{MHz}, in which the data are analysed. The excess noise at \SI{100}{MHz} is a spurious noise compatible with FM radio signals. Moreover, it is far from the frequency ranges selected for data analysis, therefore it is not relevant for the purpose of the BHD characterization.  (b) Common-Mode Rejection Ratio (CMRR) of the BHD (blue circles) and of the oscilloscope (grey circles) used for acquiring the signals. In both the cases (a,b), the laser used to test the detector is a \SI{4.72}{\micro \meter}-wavelength single-mode QCL and the incident power onto the BHD is \SI{2.2}{mW}.
  \label{fig:INPSDfreq}}
 \end{figure}

The AC signals acquired at the outputs of the two \ce{HgCdTe} detectors are digitally summed and/or subtracted. In Fig.~\ref{fig:INPSDfreq}, the Intensity Noise Power Spectral Density (INPSD) is calculated~\cite{note2} and the spectra are compared with the INPSD of the difference of the AC background signals. As described in section~\ref{sec:Theory of balanced detection}, in the linear-responsivity regime the variance of the sum and of the difference signals, measured as INPSD, provide information about the intensity noise of the incident radiation and the corresponding shot-noise level, respectively. 
\\The sum INPSD, shown in Fig.~\ref{fig:INPSDfreq}~(a) (orange trace), represents the detected intensity noise obtained using the single-mode QCL emitting at $\lambda=\SI{4.72}{\micro \meter}$, driven at \SI{712}{mA}, at a working temperature of \SI{18}{\celsius}, and after an optical attenuation of 93\%. Despite the considerable attenuation, by comparing the sum with the difference (blue trace) we can infer that the detected intensity noise of the laser is above the shot-noise level. Giving a closer look at the differential measurement in Fig.~\ref{fig:INPSDfreq}~(a) it is possible to determine the optimal working frequency range for the BHD as the interval between \SI{1}{MHz} and \SI{100}{MHz} approximately, where the difference INPSD has the typical white-noise flat trend and it is compatible with the expected ideal one-sided Power Spectral Density (PSD) of the shot noise (dashed black line). This is defined as $ PSD_{\mathrm{shot-noise}} = 2 e I$, where $e$ is the electron charge and $I$ is the detector output current~\cite{rice2016:shotnoiseinfrequency}. For high frequencies, above \SI{100}{MHz}, the data drop below the ideal shot-noise level is due to the finite bandwidth of the setup (detector and oscilloscope). The cut-off is measured as the \SI{-3}{dB} drop point of the signal, resulting in a measured bandwidth of \SI{120}{MHz}. The shot-noise-sensitivity limit of the balanced detector is given by the grey trace, which is the INPSD of the difference of the AC background signals. We have also verified that the contribution of the oscilloscope background to the electronic noise is negligible (petroleum trace). Indeed, the oscilloscope differential noise is more than \SI{20}{dB} below the detector background noise, reaching \SI{30}{dB} at \SI{30}{MHz}.

Another important parameter in the BHD characterization is the Common-mode Rejection Ratio (CMRR). The CMRR is defined as the ratio between the INPSD of the sum and the INPSD of the difference and characterizes quantitatively the noise suppression capability of the system. To measure the CMRR at different frequencies, we modulated the intensity of the laser with a square-wave signal at \SI{1}{MHz}. In this way, it is possible to test the CMRR of the BHD simultaneously at different frequencies, as the square-wave spectrum is composed of the odd harmonics of the fundamental frequency. For this characterization we used the laser in the same conditions as in Fig.~\ref{fig:INPSDfreq}~(a). For each frequency component of the square wave we have computed the CMRR as the ratio between the INPSD of the sum and that of the difference, as shown in Fig.~\ref{fig:INPSDfreq}~(b) (blue circles) \cite{note4}. The same measurements have been performed for characterizing the CMRR of the oscilloscope (grey circles) to test the instrumental limit of this measurement. In particular, we have sent the same square-wave signal, equally split, in the two channels of the oscilloscope used for the acquisition and, by measuring the INPSD of the sum and difference, we have estimated the CMRR of the oscilloscope. The analysis in Fig.~\ref{fig:INPSDfreq}~(b) is performed in the flat working region of the BHD (i.e. from \SI{1}{MHz} and \SI{100}{MHz}, as already discussed). Notice that while at lower frequencies the CMRR is limited by the oscilloscope, this is not the case for higher frequencies, where the CMRR of the BHD is lower. By the way, it is possible to find a high-frequency region around \SI{30}{MHz} where the CMRR is still over \SI{20}{dB}. 

 \begin{figure}[!htbp]
    \centering
  \includegraphics[width=0.9\columnwidth]{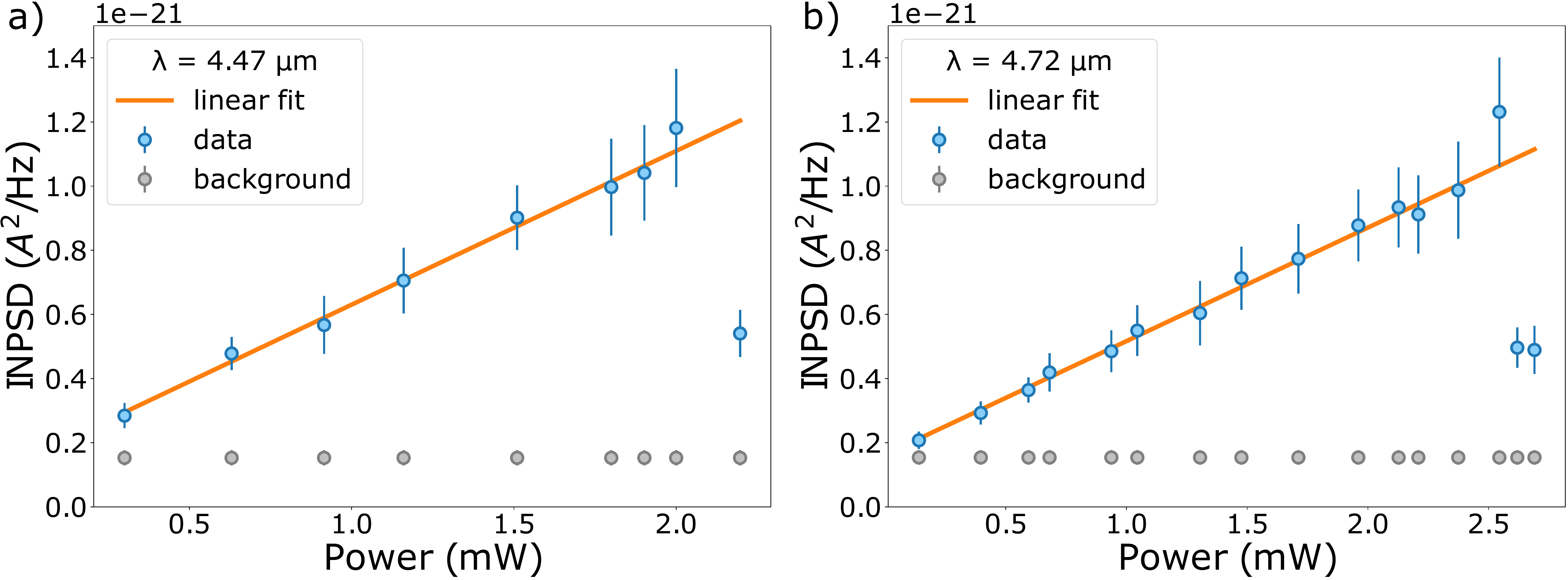}
  \caption{INPSD of the difference of the AC output signals versus the incident power of the radiation. Each point corresponds to the average level in the selected frequency window of \SI{3}{MHz} centred at \SI{30}{MHz}. The lasers used to test the detectors are two single-mode QCLs, emitting at \SI{4.47}{\micro \meter} (a) and \SI{4.72}{\micro \meter} (b), in fixed working conditions (temperature and current). The incident power is changed via the variable attenuator depicted in Fig.~\ref{fig:50-50setup}.} 
  \label{fig:INPSDvsP}
\end{figure}
In general, to minimize the effect of the background on the measured INPSD it is convenient to identify the optimal working region of the BHD, where the shot noise of the incident radiation is well above the background noise and the responsivity is kept linear at the same time. Indeed, only in the linear-responsivity regime of the detectors, current fluctuations are directly proportional to the photon-number fluctuations. Therefore, by measuring the statistics of the current, it is possible to obtain direct information on the statistics of the light under investigation. Furthermore, this is the optimal working region for quantum application as well. Indeed, it is exactly in the range where the shot noise is well above the background noise that it is possible to unveil sub-shot-noise fluctuations, expected for quantum states such as intensity-squeezed states~\cite{loudon:2000quantum}. Given these considerations, we selected a \SI{3}{MHz} frequency window, centred at \SI{30}{MHz}, for the spectral analysis. Here, the sum signal is less perturbed by classical and technical noise contributions and the incident radiation noise is indeed closer to the shot-nose level and more compatible with a coherent state. At the same time, this frequency window is far enough from the high-frequency cut-off of the detectors. By computing the average level of the spectra in this window, we verified that the INPSD of the difference shows a linear trend with the LO incident optical power and that the BHD is shot-noise limited. This demonstrates that our detector is suitable for measuring the fluctuations of the incident radiation at the shot-noise level and below. As shown in Fig.~\ref{fig:INPSDvsP}, at \SI{30}{MHz} the measured differential INPSDs are up to 7 (6) times above the background level for $\lambda = \SI{4.47}{\micro \meter}$ (\SI{4.72}{\micro \meter}), i.~e. the maximum clearance of the BHD is 7 (6). This leads to an equivalent optical efficiency, as defined in~\cite{appel2007electronic}, of 86\% for an incident radiation at a wavelength of \SI{4.47}{\micro \meter} (83\% at \SI{4.72}{\micro \meter}). By multiplying this value for the measured quantum efficiency (41\% at $\lambda=\SI{4.47}{\micro \meter}$), we find an effective quantum efficiency of 35\%. To achieve the best performance in terms of clearance and linear responsivity, the incident power of the LO has to be accurately selected. According to our characterization, the optimal LO power impinging on the beam splitter is \SI{2.0}{mW} at \SI{4.47}{\micro \meter} and \SI{2.5}{mW} at \SI{4.72}{\micro \meter}. Notice that in the characterization at \SI{4.47}{\micro \meter} (Fig.~\ref{fig:INPSDvsP}) saturation is achieved just above  \SI{2.0}{mW} incident power, significantly lower than the saturation level measured for \SI{4.72}{\micro \meter} ($P>\SI{2.5}{mW}$). This is in agreement with the responsivity peak of our HgCdTe detectors, which is centred around \SI{4.5}{\micro \meter}.
It is also important to notice that the saturation level depends also on the transimpedance amplification system: proper adjustments of the electronic amplification can allow higher LO power detection while preventing early-power saturation.
\begin{figure}[!htbp]
    \centering
  \includegraphics[width=0.95\columnwidth]{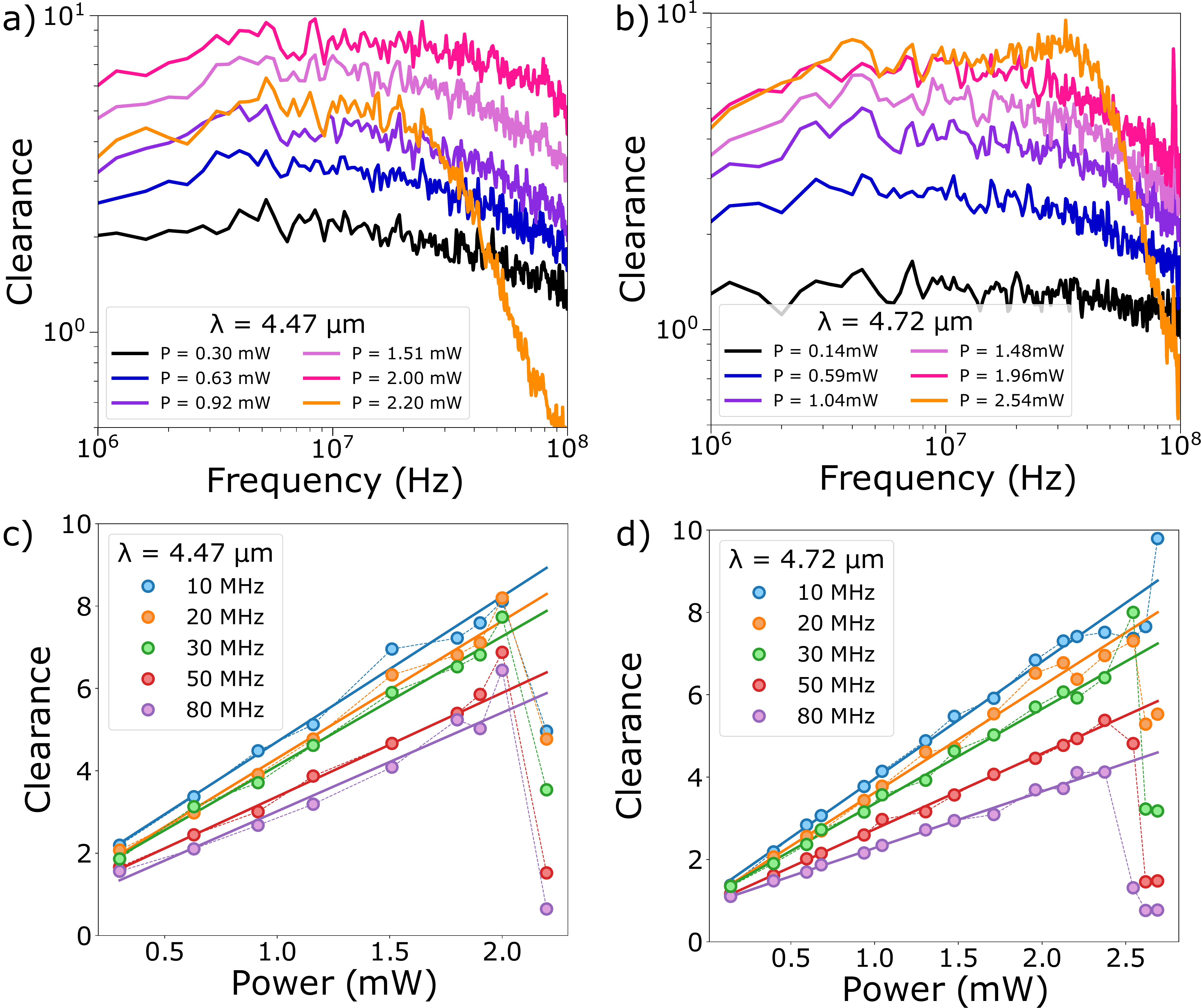}
  \caption{ Clearance measured at \SI{4.47}{\micro m} (a) and \SI{4.72}{\micro m} (b) as a function of the Fourier frequency for several values of incident power on the BHD. The clearance spectra at \SI{4.47}{\micro m} (c) and at \SI{4.72}{\micro m} (d) have been integrated over a frequency window of \SI{3}{MHz} centered at different values, as a function of the incident power. Graphs (c,d) show experimental data (circles) and  linear fits (lines).
  \label{fig:Clearance}}
\end{figure}

A thorough analysis of the performance of the BHD in terms of clearance and linearity is shown in Fig.~\ref{fig:Clearance}. In particular, the clearance spectra (a,b), for several incident power values and for both the wavelengths used, are plotted in the \SIrange{1}{100}{MHz}-frequency region, i.e. where the differential spectrum is flat, as evidenced in the previous discussion (Fig.~\ref{fig:INPSDfreq}~(a)). In particular, Fig.~\ref{fig:Clearance}~(a) and (b) show two different steps of saturation regime: in graph (a) the clearance at \SI{2.20}{mW} (orange trace) is characterized by a saturation visible in the whole spectrum, while in graph (b) the curve corresponding to \SI{2.54}{mW} is at the edge of power saturation, clearly visible only for high frequencies (>\SI{40}{MHz}). According to the analysis shown in Fig.~\ref{fig:INPSDvsP}, in Fig.~\ref{fig:Clearance} the spectra exhibit a higher clearance before saturation at \SI{4.47}{\micro \meter}, reaching a value up to 8 and, consequently, an effective quantum efficiency up to 36\%. To evidence the linearity of the clearance with the incident power of the BHD, we have integrated the spectra in a \SI{3}{MHz} window centered at different frequencies, as reported in Fig.~\ref{fig:Clearance}~(c) and (d). In Fig.~\ref{fig:Clearance}~(c) it is clear that for an incident power of \SI{2.0}{mW} the clearance deviates from the linear trend for frequencies above \SI{10}{MHz}, suggesting a beginning of saturation. In summary, from Fig.~\ref{fig:Clearance}~(c) and (d) it is possible to conclude that the detector shows a linear behavior with the incident power until the saturation level is achieved, for different Fourier frequencies going from \SIrange{10}{80}{MHz}. In addition, the graphs show that the clearance decreases by increasing the central frequency of such analysis. 

The achieved values of overall detection efficiency and clearance of our BHD are very encouraging. With the present values, our BHD will be able to detect quantum states of light in the MIR such as squeezed states. Quantum characterization of single-photon states as well as more exotic quantum states is also possible, although quantum state tomography requires quantum efficiencies above 50\% to retrieve negative-valued Wigner functions \cite{lvovsky2009continuous}. However, more sophisticated criteria can be applied to certify non-classicality of the light under investigation when the quantum efficiency is lower than 50\% \cite{biagi:2021}. 
Instead, for quantum information processing such as continuous-variables quantum teleportation \cite{paris:2003} and for long-distance continuous-variable quantum communications in free space (e.g. satellite-based \cite{Dequal2021}), the overall detection efficiency needs to be improved. 

\section{Conclusion}

We have presented a mid-infrared balanced detection system suitable for quantum characterization of light via homodyne detection. In particular, we have proven its capability to achieve shot-noise-limited detection, by showing that the differential signal retrieved at the output of the beam splitter is directly proportional to the incident power. The main features of the setup are \SI{120}{MHz} bandwidth, quantum efficiency up to 41\%, saturation for incident power higher than \SI{2.0}{mW} at the peak-responsivity wavelength of the HgCdTe detectors, and 50/50 DC splitting ratio with 0.2\% uncertainty. In this work, the wavelength dependence of the BHD responsivity and saturation is studied by testing the setup with two QCLs emitting at \SI{4.72}{\micro \meter} and \SI{4.47}{\micro \meter}. The spectral analysis of the clearance operated for different values of incident power on the BHD and at different FFT frequencies evidences that the maximum clearance, while keeping a linear response and avoiding saturation, is up to 8, leading to an effective quantum efficiency of 36\%. This value is achievable at a Fourier frequency of \SI{10}{MHz}, that is an optimal working region also for the CMRR achieving over \SI{30}{dB}. The CMRR analysis shows a significant frequency dependence in the noise extinction in the presented setup: the CMRR decreases as the frequency increases. In general, for balanced homodyne detection applications, the optimal working region for the LO is where the clearance is maximum. Indeed, this is the optimal range where sub-shot-noise fluctuations can be observed. Furthermore, by exploiting the possibilities given by the digital mathematical operations, the BHD can easily be adapted to measure not only sum and difference but also product or even more complex mathematical operations. This makes our BHD a versatile tool suitable not only for homodyne detection but also for other schemes, e.~g. second-order correlation measurements, that require a mathematical operation on split optical beams. This detector can be used for classical measurements as well, taking all the advantages of a 50/50 balanced detection where the common noise contributions from the two balanced signals are subtracted at the shot-noise level. Given its versatility, this detector represents an important step towards the quantum characterization of mid-infrared light. The first test results reported here show that such a scheme can represent a novel setup for quantum characterization of mid-infrared light, suitable for demonstrating quantum-state generation in mid-infrared sources.
\section*{Acknowledgments}
The authors gratefully thank the collaborators within the consortium of the Qombs Project: Prof. Dr. Jérome Faist (ETH Zurich) for having provided the quantum cascade lasers and the company ppqSense for having provided the ultra-low-noise laser current drivers (QubeCLs). \\ 
The Authors acknowledge financial support from the European Union’s Horizon 2020 Research and Innovation Programme (Qombs Project, FET Flagship on Quantum Technologies grant n. 820419; Laserlab-Europe Project grant n. 871124) and from the Italian ESFRI Roadmap (Extreme Light Infrastructure - ELI Project). 
\section*{Disclosures}
The authors declare no conflicts of interest. 
\section*{Data availability}
The data that support the findings of this study are available from the corresponding author upon reasonable request.

\bibliographystyle{ieeetr_CA}
\bibliography{references}

\begin{thebibliography}{10}
\bibitem{note3}
The complete VIGO detectors' code is PVI-4TE-5-2x2-TO8-wAl203-36+MIP-10-250M. A
  DC-output port has been added in house. As reported on the producer
  datasheet, the detectors are characterized by a declared time constant $\tau
  \leq \SI{80}{ns}$, a transimpedance of \SI{1.95E+4}{V/A} and an output noise
  density (averaged over \SI{1}{MHz} to high cut-off frequency of
  \SI{180}{MHz}) of $\SI{400}{nV/ \sqrt{Hz}}$. These last two quantities are
  measured on a load $R_{\mathrm{LOAD}} = \SI{50}{\Omega}$. Notice that the
  data shown in this article have been acquired using the oscilloscope in
  high-input-resistance mode with $R_{\mathrm{LOAD}} = \SI{1}{M\Omega}$.
  Additional information on the detectors is available in the datasheets on the
  producer website.
\bibitem{note2}
The INPSDs are calculated as twice the value obtained via the matplotlib Python
  library function matplotlib.mlab.psd(). We apply the hanning windowing and
  normalize the data to the noise bandwidth. The NFFT parameter, i.e. the
  number of data points used in each block for the FFT, is 1/50 of the length
  of the input data. The input data are the AC current signals scaled by the AC
  gain.
\bibitem{note4}
To correctly evaluate the CMRR at each frequency component of the square-wave
  signal, we have set the modulation signal amplitude to a sufficiently high
  value ensuring that the residual noise component emerges from the noise after
  the subtraction.
\end{thebibliography}

\end{document}